%% file: main.tex
\documentclass[conference]{IEEEtran}
\IEEEoverridecommandlockouts

\usepackage{cite}
\usepackage{amsmath,amssymb,amsfonts}
\usepackage{algorithm, algorithmic}
\usepackage{graphicx}
\usepackage{textcomp}
\usepackage{xcolor}
\usepackage{listings}

\lstdefinestyle{mystyle}{
    commentstyle=\color{codegreen},
    keywordstyle=\color{magenta},
    stringstyle=\color{codepurple},
    basicstyle=\ttfamily\footnotesize,
    breakatwhitespace=false,         
    breaklines=true,                 
    captionpos=b,                    
    keepspaces=true,                 
    numbers=left,                    
    numbersep=5pt,                  
    showspaces=false,                
    showstringspaces=false,
    showtabs=false,                  
    tabsize=2,
    xleftmargin=2em,
    frame=single
}

\newcommand{\toolname}{Rampo}

\newtheorem{theorem}{\textbf{Theorem}}[section]
 
\newtheorem{problem}[theorem]{\textbf{Problem}}
\newtheorem{definition}[theorem]{\textbf{Definition}}

\usepackage{soul}

\renewcommand{\baselinestretch}{0.975}

\begin{document}

\input{sections/1_title_abst}

\input{sections/2_introduction}
\input{sections/3_related_work}
\input{sections/4_framework}
\input{sections/6_experiment_result}
\input{sections/7_conclusion}

\bibliographystyle{IEEEtran}
\input{main.bbl}

\end{document}

%% file: sections/1_title_abst.tex
\title{
\toolname: 
A CEGAR-based Integration of Binary Code Analysis and System Falsification for Cyber-Kinetic Vulnerability Detection
}

\author{\IEEEauthorblockN{Kohei Tsujio, Mohammad Abdullah Al Faruque, Yasser Shoukry} \
\IEEEauthorblockA{\textit{Department of Electrical Engineering and Computer Science} \\
\textit{University of California, Irvine}}
}

\maketitle

\begin{abstract}
    Cyber-physical systems (CPS) play a pivotal role in modern critical infrastructure, spanning sectors such as energy, transportation, healthcare, and manufacturing. These systems combine digital and physical elements, making them susceptible to a new class of threats known as \emph{cyber kinetic} vulnerabilities. Such vulnerabilities can exploit weaknesses in the cyber world to force physical consequences and pose significant risks to both human safety and infrastructure integrity. This paper presents a novel tool, named \toolname, 
    that can perform binary code analysis to identify cyber kinetic vulnerabilities in CPS. The proposed tool takes as input a Signal Temporal Logic (STL) formula that describes the kinetic effect---i.e., the behavior of the ``physical'' system---that one wants to avoid. The tool then searches the possible ``cyber'' trajectories in the binary code that may lead to such ``physical'' behavior. This search integrates binary code analysis tools and hybrid systems falsification tools using a Counter-Example Guided Abstraction Refinement (CEGAR) approach. In particular, \toolname  ~starts by analyzing the binary code to extract symbolic constraints that represent the different paths in the code. These symbolic constraints are then passed to a Satisfiability Modulo Theories (SMT) solver to extract the range of control signals that can be produced by each of the paths in the code. The next step is to search over possible ``physical'' trajectories using a hybrid systems falsification tool that adheres to the behavior of the ``cyber'' paths and yet leads to violations of the STL formula. Since the number of ``cyber'' paths that need to be explored increases exponentially with the length of ``physical'' trajectories, we iteratively perform refinement of the ``cyber'' path constraints based on the previous falsification result and traverse the abstract path tree obtained from the control program to explore the search space of the system.
    To illustrate the practical utility of binary code analysis in identifying cyber kinetic vulnerabilities, we present case studies from diverse CPS domains, showcasing how they can be discovered in their control programs. In particular, compared to off-the-shelf tools, our tool could compute the same number of vulnerabilities while leading to a speedup that ranges from $3\times$ to $98\times$.
\end{abstract}

%% file: sections/2_introduction.tex
\section{Introduction}
\label{sec:introduction}

In an era where the backbone of modern society—power grids, transportation networks, healthcare facilities, and industrial control systems—rely extensively on cyber-physical systems (CPS), understanding and protecting against vulnerabilities has never been more critical. In particular, the convergence of physical and digital systems in critical CPS infrastructure has introduced new dimensions of risk, giving rise to a class of threats known as cyber kinetic vulnerabilities~\cite{theDawnofCyberKinetic, AlFaruqueDAC,7827651}. These vulnerabilities represent a unique breed of cyber-physical security challenges, where the exploitation of software vulnerabilities can lead to potentially catastrophic physical consequences. That is, cyber kinetic vulnerabilities blur the lines between digital and physical security, requiring a holistic approach that can analyze software controlling physical processes from the point-of-view of the induced physical or kinetic behavior.

Unfortunately, there is currently a disconnect within the field of CPS formal verification and analysis. On the one side, the celebrated achievements in the software and hardware verification field~\cite{symbolic_model_checking, kroening2014cbmc, smt_model_checking,holzmann1997model,boulanger2013industrial} are inadequate for analyzing the dynamics of the underlying physical components of CPS. On the other hand, the techniques developed in the control theory community have focused mainly on the formal verification of the dynamics of physical/mechanical systems~\cite{tabuada2009verification,chen2013flow,bansal2017hamilton,s-taliro,donze2010breach} while ignoring the complexity of the computational and cyber components. This paper aims to provide a comprehensive framework for tackling cyber kinetic vulnerabilities, one that not only identifies potential threats but also evaluates their physical consequences. In a rapidly evolving digital landscape, this integrated approach is crucial for protecting critical infrastructure and ensuring the safety and reliability of essential services for society at large.

This paper delves into the compelling realm of binary code analysis and falsification as an indispensable facet of safeguarding CPS against cyber kinetic vulnerabilities. Binary code, the low-level representation of software, is the critical interface where the virtual and physical worlds intersect. By subjecting binary code to rigorous scrutiny, researchers and security professionals can unearth vulnerabilities that are invisible at higher abstraction levels and thus lay the foundation for enhancing the resilience of critical infrastructure. In particular, compared to code-level, binary code analysis provides several advantages, including (i)~\emph{Visibility into Compiled Code:} Binary code analysis involves examining the compiled form of a software application, providing insights into the actual instructions executed by the processor. This level of analysis is essential for understanding how the software interacts with the hardware, as it focuses on the executable, machine-readable code (ii)~\emph{Reverse Engineering:} Reverse engineering binary code can reveal hidden or obfuscated vulnerabilities, which might not be readily apparent at the source code level. It is a valuable technique for understanding proprietary or closed-source software, and (iii)~\emph{Legacy Systems:} In several CPS applications, only the binary code is available, especially for older or proprietary software. Binary code analysis is the primary method for assessing the security of such systems.

\begin{figure}[!t]
    \centering
    \includegraphics[width=\linewidth]{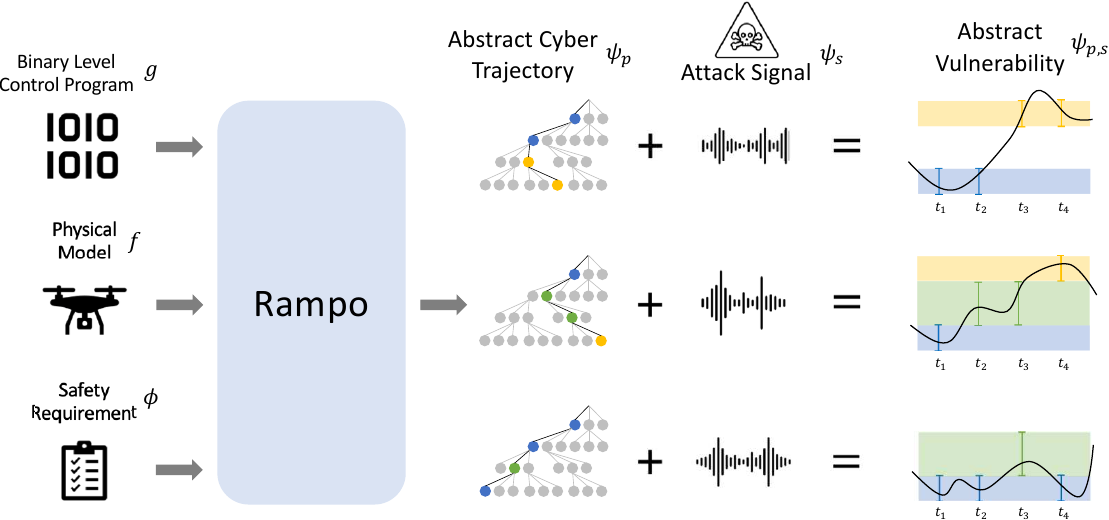}
    \caption{\toolname \ takes as input the binary code of the control program, a black-box model that represents the physics of the system, and cyber-kinetic safety requirements captured in STL. The outputs are the cyber trajectories inside the binary code and the corresponding physical system behavior that can lead to violation of the STL requirements. Moreover, \toolname \ can also output attacks on the system's sensors that can also lead to kinetic vulnerabilities.}
    \label{fig:input_output}
\end{figure}

This paper introduces a novel tool called \toolname.
As shown in Figure~\ref{fig:input_output}, \toolname~takes as an input a Signal Temporal Logic (STL) formula that describes the kinetic effects one would like to avoid. For example, this STL formula can encode the requirement that the voltages and currents of the electric distribution buses should be within acceptable safe ranges, drones should approach landing sites with specific velocity profiles to avoid a crash, or the machinery in the industrial control systems should operate within safe operation regimes. The objective of our tool is to simultaneously identify vulnerabilities in the binary code used to control this CPS along with physical attacks (e.g., changes in the sensor measurements) that can lead to a violation of the STL formula and, hence, an undesired kinetic vulnerability. In other words, \toolname \ is designed to simultaneously analyze the binary code and the dynamics of the physical system to identify vulnerabilities in both the binary code and the physical system that can lead to kinetic consequences.

The design of a tool that can simultaneously analyze the binary code and the dynamical behavior of physical systems within CPS faces two significant complexity challenges. First, performing such analysis necessitates tools that can reason about constraints defined over discrete and real variables, respectively. Unfortunately, tools like Satisfiability Modulo Theories (SMT) solvers and Mixed-Integer Programming (MIP) do not scale well for complex systems. Second, due to the dynamic behavior of CPS, one needs to analyze temporal trajectories over both the cyber and physical systems. That is, while existing binary code analysis focuses on analyzing all the possible paths inside the code within one execution of the code, in CPS, the control code is executed periodically where the path taken within the same binary code may differ from one time to another depending on the state of the physical system. Hence, consider, for the sake of an example, a simple code with three different possible paths (corresponding to different if-else conditions in the code). A classical binary code analysis will aim to cover all these three paths. Nevertheless, considering temporal trajectories of 10-time steps in a CPS application, one must consider all $3^{10} = 59049$ possible paths that may arise from computing the control signal 10 times within a trajectory, a daunting computational problem. 

\begin{figure}[t]
    \centering
    \includegraphics[width=\linewidth]{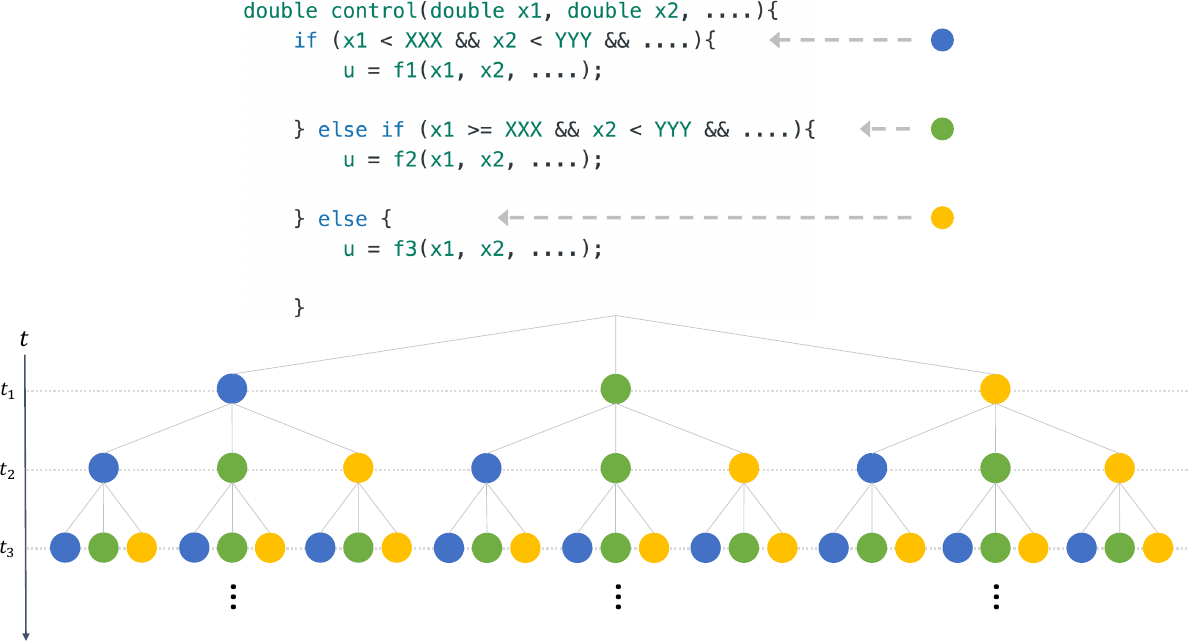}
    \caption{An example of a code with three paths (marked with blue, green, and yellow). When considering the temporal evolution of a CPS, one needs to take into account the different combinations of these three paths across time, which leads to an exponential increase in the overall number of paths.}
    \label{fig:tree}
\end{figure}

To solve the intertwined challenges mentioned above, \toolname \ integrates two state-of-the-art techniques from binary analysis and hybrid systems falsification into a unified framework using a Counter-Example Guided Abstraction Refinement (CEGAR) approach~\cite{cegar}. In particular, we use binary analysis tools to perform symbolic execution over the binary code to extract SMT constraints representing different paths inside the code. Using SMT solvers, we solve an optimization problem to identify the bounds on the control signal that can be generated from every possible path in the binary code. To avoid the combinatorial explosion resulting from considering different paths along a temporal trajectory of the CPS, we \emph{abstract} several possibilities of code path executions and use this abstraction to \emph{guide} a falsification of the physical system against the STL requirements. Falsification tools use stochastic optimization algorithms to find trajectories. If a trajectory of the physical system is found, we refine our abstraction using the information extracted from this trajectory (also referred to as an abstract counter-example). By refining the abstraction iteratively, we can ensure the coverage of all possible paths inside the binary code along the temporal evolution of the CPS. 
In summary, this paper introduces several novel contributions summarized as follows:
\begin{itemize}
    \item We propose a novel framework to identify cyber-kinetic and cyber-physical-kinetic vulnerabilities in software used to control physical systems.
    \item We propose a Counter-Example Guided Abstraction Refinement approach to harness the exponential growth in the search space.
    \item We provide an ablation study that shows the benefits of the proposed framework, both in terms of the effectiveness of finding vulnerabilities and in terms of scaling more favorably to off-the-shelf tools.
\end{itemize}

%% file: sections/3_related_work.tex
\section{Related works}
\label{sec:related_works}
Analyzing CPS software for correctness and safety has been an active area in the past few decades. For example, Fuzz testing has become an instrumental technique in the domains of software verification and security analysis. Several works address the fuzz testing of CPS software for automatic test generation~\cite{sheikhi2022coverage} and for the discovery of security vulnerabilities~\cite{serpanos2021fuzzing,moukahal2021vulnerability,fowler2018fuzz}. Unfortunately, the work in\cite{serpanos2021fuzzing,moukahal2021vulnerability,fowler2018fuzz} (and others) have focused either on simple physical systems or vulnerabilities with no kinetic consequences.

Another prominent technique for identifying vulnerabilities in software is through symbolic execution techniques. \textit{KLEE}~\cite{KLEE} and \textit{angr}~\cite{angr} are some of the most powerful analysis tools. By utilizing symbolic execution and SMT solvers, these tools can extract path information in a binary code and find several types of vulnerabilities. Unfortunately, these techniques focus mostly on traditional cyber vulnerabilities and do not generalize to reason about cyber-kinetic vulnerabilities due to their lack of analyzing models of physical systems.

One of the widely applied techniques to discover bugs in CPS is
Falsification~\cite{9315067,7886680,s-taliro}. While traditional falsification engines do not guarantee to cover all possible paths inside the control software, new directions in this field have shown new techniques to provide such code coverage~\cite{georgios, 7318257}. Unlike the work in~\cite{georgios, 7318257}, this paper focuses on coverage of code paths \emph{along temporal evolution of the code}. As motivated in Figure~\ref{fig:tree}, the temporal evolution of the code leads to an exponential growth in the space of all possible paths that a code can take over time. This challenge of dealing with the temporal evolution of code (or as we call it cyber trajectories), is one of the main motivations behind our CEGAR-based approach.

%% file: sections/4_framework.tex
\section{Problem Definition}
\label{sec:problem}

\subsection{Notation}
We use the symbols $\mathbb{N}$, $\mathbb{R}$, and $\mathbb{B}$ to denote the set of natural, real, and Boolean numbers, respectively. Additionally, we use the symbol $\mathbb{FP}$ to denote the set of Floating Point numbers. For ease of notation, we do not distinguish between 32-bit (single) or 64-bit (double) floating point numbers. We use $\land$, $\lor$, and $\lnot$ to represent the logical AND, OR, and NOT operators, respectively. Given a sequence $a = \{a_t\}_{t = 0}^N$, we denote by $\texttt{Element}(a,i)$ the $i$th element of the sequence $a$.

\subsection{Cyber-Kinetic Vulnerabilities}

We consider nonlinear, discrete-time dynamical systems:
\begin{align}
    x_{t+1} &= f(x_t, u_t), \quad
    y_t = h(x_t), \quad
    u_t = g(y_t),
\end{align}
where $x_t \in \mathbb{R}^n$ is the state of the system at time $t \in \mathbb{N}$, $u_t \in \mathbb{FP}^{m} \subset \mathbb{R}^m$ is the action at time $t$, and $y_t \in \mathbb{FP}^o \subset \mathbb{R}^o$ is the sensor (output) measurements at time $t$. As a cyber-physical system, we assume that the system is controlled by a control program $g:\mathbb{FP}^o \rightarrow \mathbb{FP}^m$. Without loss of generality, any program can be decomposed using the so-called path constraints~\cite{de2021symbolic} as:
\begin{align*}
    g(y) = & [c^{(1)}(y) \Rightarrow u = g^{(1)}(y)] \\
    \land & [c^{(2)}(y) \Rightarrow u = g^{(2)}(y)] \\
    & \qquad \vdots \\
    \land & [c^{(k)}(y) \Rightarrow u = g^{(k)}(y)],
\end{align*}
where $c^{(i)}:\mathbb{FP}^o \rightarrow \mathbb{B}$ is the $i$th path constraint of the program $g$ and $g^{(i)}:\mathbb{FP}^o \rightarrow \mathbb{FP}^m$ is the path function, i.e., the function that is executed at the $i$th path of $g$. 

\noindent \textbf{Example:}
Consider the following control program $g$:
\begin{lstlisting}
double control(double y1, double y2){
  if (-1 <= y1 && y1 <= -0.5){
    if (y2 > 0){
        u = 4 * y1 - 6;
    }else{
        u = 0;
    }
  }else{ 
    u = -4 * y1 - 6;
  }
  return u;
}
\end{lstlisting}
This control program consists of $3$ paths with the following path constraints:
\begin{align*}
    c^{(1)}(y) &= (-1 \le y1) \land (y1 \le -0.5) \land (y2 > 0), \\
    c^{(2)}(y) &= (-1 \le y1) \land (y1 \le -0.5) \land (y2 \le 0), \\
    c^{(3)}(y) &= \lnot \left((-1 \le y1) \land (y1 \le -0.5)\right).
\end{align*}
The corresponding path functions are:
\begin{align*}
    g^{(1)}(y) &= 4y1 - 6, \quad g^{(2)}(y) = 0, \quad g^{(3)}(y) = -4 y1 - 6.
\end{align*}
It is important to note that we do not assume prior knowledge of the path constraints and the corresponding path functions, as they will be extracted automatically from the binary code representing the control program.

Given a control program $g$ that consists of $k$ paths, we denote by $\texttt{PATH}:\mathbb{FP}^o\rightarrow \{1, \ldots, k\}$ the function that returns the path index of an input $y$. That is:
\begin{align}
    \texttt{PATH}(y) = p \Longleftrightarrow c^{(p)}(y) = \texttt{True}.
\end{align}
Note that the function $\texttt{PATH}$ is well defined since path constraints are mutually exclusive, and hence, an input $y$ can only be processed by only one path inside the code.

Using this notation, we can now define the cyber trajectory of a system as the index of the control program's path taken at different time instances as follows.

\begin{definition}[Cyber Trajectories]
Given a control program $g$ with $k$ paths, a sequence $\psi_p = \{p_t\}_{t = 0}^H$, with $p_t \in \{1, \ldots, k\}$, is called a cyber trajectory of the physical system $f$ if there exists a corresponding trajectory of the physical system $\psi_x = \{x_t\}_{t = 0}^H$ with $x_{t+1} = f(x_t, g(h(x_t))) \ (t < H)$ such that $p_t = \texttt{PATH}(h(x_t))$ for all $t \in \{0, \ldots, H\}$.

\end{definition}

We are interested in enumerating all cyber trajectories that can lead to physical system trajectories that violate a safety requirement $\varphi$. In this paper, we assume that the safety requirement $\varphi$ is captured by a Signal Temporal Logic (STL) formula. 
For the formal definition of STL syntax and semantics, we refer the reader to~\cite{maler2004monitoring}. Using this notation, the problem of interest can be defined as follows. 

\begin{problem}[Enumeration of Cyber-Kinetic Vulnerabilities]
Given an STL formula $\varphi$, a controller program $g$ with $k$ paths, and a physical system model $f$. Find the set $\Psi_p^\varphi$ of cyber trajectories that may lead to cyber-kinetic vulnerabilities, i.e., $\Psi_p^\varphi$ is defined as:
\begin{align}
\Psi_p^\varphi = &\Big\{ \{p_t\}_{t = 0}^H \in \{1,\ldots,k\}^{H+1} \; \vert \;  \nonumber \\
& \exists \psi_x = \{x_t\}_{t = 0}^H . \big[ 
x_{t+1} = f(x_t, g(h(x_t))) \ (t < H), \nonumber\\ 
& \psi_x \not \models \varphi, \; p_t = \texttt{PATH}(h(x_t)), \forall t \in \{0, \ldots, H\} \big]
\Big\}.
\end{align}
\vspace{-3mm}
\label{prob:cyber_kinetic}
\end{problem}

In other words, a cyber trajectory $\psi_p$ is considered a cyber-kinetic vulnerability if there exists an associated trajectory of the physical system $\psi_x$ that violates the safety requirement $\varphi$.

\begin{figure*}[!t]
    \centering
    \includegraphics[scale=0.8]{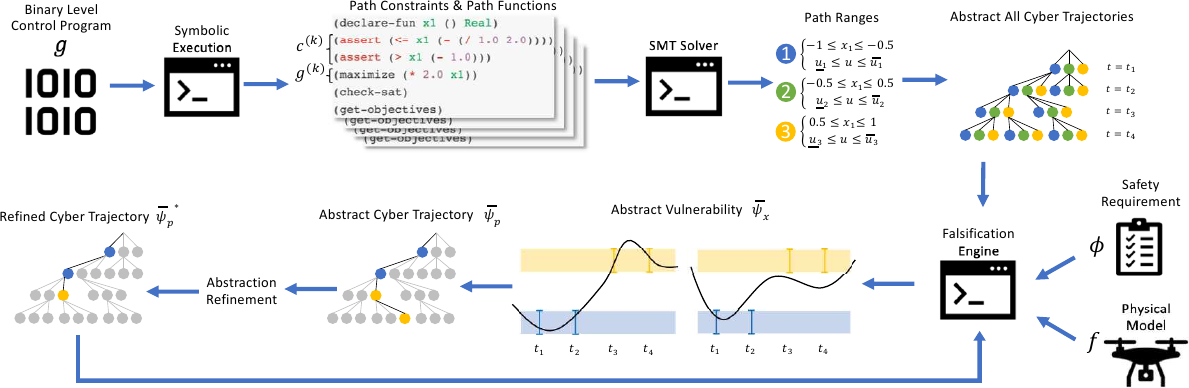}
    \vspace{-3mm}
    \caption{A pictorial representation of the proposed framework. Binary-level control programs are analyzed using symbolic execution tools to extract the path constraints, path functions, and the range of the control signals associated with each path in the program. Next, a Counter-Example Guided Abstraction Refinement (CEGAR) is used to abstract all cyber trajectories for a horizon $H$, and a falsification engine is used to search for trajectories of the physical system that violate the safety requirements. The cyber trajectories are then refined around the falsifying trajectories until all concrete cyber-kinetic vulnerabilities are found.}
    \label{fig:framework_overview}
\end{figure*}

\subsection{Cyber-Physical-Kinetic Vulnerabilities}
While Problem~\ref{prob:cyber_kinetic} focuses on finding vulnerabilities in the controller program that lead to violation of safety requirements, we are also interested in generalizing this problem into scenarios when an attacker is capable of manipulating the physical sensor measurements $h(x_t)$ before it is processed by the control program $g$~\cite{shoukry2013non,shoukry2015pycra,shoukry2017secure,shoukry2015event,barua2020hall,barua2022wolf}. Typically, these sensor manipulations are of relatively small magnitude $\overline{s}$ compared to the sensor noise levels to avoid being detected. That is, we consider the system:

\begin{align}
    x_{t+1} &= f(x_t, u_t), \quad
    y_t = h(x_t) + s_t, \quad
    u_t = g(y_t)
\end{align}
where $s_t$ is the physical attack signal at time $t$ with $\Vert s_t \Vert \le \overline{s}$. In such a setup, given a trajectory of the physical system $\psi_x$ and a trajectory of the physical attack signal $\psi_s = \{s_t\}_{t = 0}^H$, the corresponding cyber trajectory $\psi_p$ is defined as $\psi_p = \{ \texttt{PATH}(h(x_t) + s_t) \}_{t = 0}^H$.

\begin{problem}[Enumeration of Cyber-Physical-Kinetic Vulnerabilities]
Given an STL formula $\varphi$, a controller program $g$ with $k$ paths, a physical system model $f$, and a limit on the physical attack signal $\overline{s}$. Find the set $\Psi_{p,s}^\varphi$ of cyber-physical-kinetic vulnerabilities defined as:
\begin{align}
\Psi_{p,s}^\varphi = \Big\{ &\{(p_t, s_t)\}_{t = 0}^H \in \{1,\ldots,k\}^{H+1} \times \mathbb{FP}^{H+1}\; \vert \; \nonumber \\
& \exists \psi_x = \{x_t\}_{t = 0}^H . \big[ 
x_{t+1} = f(x_t, g(h(x_t) + s_t)) \ (t < H),\nonumber \\
&\psi_x \not \models \varphi, \;
p_t = \texttt{PATH}(h(x_t) + s_t), \nonumber \\
& \Vert s_t \Vert \le \overline{s}, \forall t \in \{0, \ldots, H\} \big]
\Big\}.
\end{align}

\label{prob:CPS_kinetic}
\end{problem}

\section{Framework}
\label{sec:framework}

\subsection{Overview of the \toolname~Framework}
As pictorially shown in Figure~\ref{fig:framework_overview}, given a binary-level control program $g$, our first step is to extract the path constraints $c^{(1)}, \ldots, c^{(k)}$ from $g$. To this end, we use off-the-shelf symbolic code execution engines that can extract SMT constraints representing different paths in the control program (Line~\ref{ln:symbolic} in Algorithm~\ref{alg:rampo}). 

The next step is to analyze the extracted paths to find cyber trajectories that can lead to cyber-kinetic vulnerabilities. Nevertheless, and as motivated in Figure~\ref{fig:tree}, the number of cyber trajectories increases exponentially as a function of the number of paths $k$ and the length of the trajectory $H$. In particular, for trajectories of length $H$ and a control program with $k$ paths, our tool is now challenged with analyzing all $k^H$ possible cyber trajectories. To harness the combinatorial explosion, we resort to an iterative Counter-Example Guided Abstraction Refinement (CEGAR) process. In this process, cyber trajectories go through several levels of abstraction that will be iteratively refined later in the process to ensure coverage of all possible cyber trajectories.

The first level of abstraction is to consider the ``range'' of the control signal $u$ that each program path can produce. Without loss of generality, and for the sake of notation, we will assume that control signal $u$ is scalar --- nevertheless, our framework is designed to support multidimensional control signals. For such a case, the maximum and minimum of the control signal $u$ within the $i$th path $p$ of $g$, denoted by $\overline{u}_p$ and $\underline{u}_p$, can be computed by solving the following optimization problem:
\begin{align}
    \overline{u}_p = & \max_{y \in \mathbb{PF}^o} g^{(p)}(y) \quad \text{ subject to } \quad c^{(p)}(y) = \texttt{True},\\
    \underline{u}_p = & \min_{y \in \mathbb{PF}^o} g^{(p)}(y) \quad \text{ subject to } \quad c^{(p)}(y) = \texttt{True}.
\end{align}
Thanks to the current advances in SMT solvers (e.g., Z3~\cite{z3}), one can efficiently solve the optimization problem above to obtain the \emph{path ranges} $R = \{(\overline{u}_1, \underline{u}_1), \ldots, (\overline{u}_k, \underline{u}_k)\}$. Given these path ranges, we can now abstract a cyber trajectory using the ranges of control signals at each time step. That is, for any cyber trajectory $\psi_p = \{p_t\}_{t = 0}^H$, the corresponding ``range'' of control signals within this trajectory is defined as:
\begin{align}
    \overline{\texttt{Range}}(\psi_p) = \Big\{ \overline{u}_{p_t} \Big\}_{t = 0}^H, \quad \underline{\texttt{Range}}(\psi_p) = \Big\{ \underline{u}_{p_t} \Big\}_{t = 0}^H 
\end{align}

The second level of abstraction combines multiple cyber trajectories into an ``abstract cyber trajectory''. Such abstraction is done by combining the ranges of control signals across different cyber trajectories. 

In particular, the first iteration of \toolname~abstracts all cyber trajectories into one. The ranges of control signals along all cyber trajectories are used to augment the STL-based safety requirements $\varphi$ as follows:
\begin{align}
    \lnot \phi = \lnot \varphi \land \left[ \bigwedge_{t = 0}^H \min_{p \in \{1,\ldots,k\}} \underline{u}_p \le u_t \le \max_{p \in \{1,\ldots,k\}} \overline{u}_p \right].
\label{eq:phi_1}    
\end{align}
The logical negation $\lnot$ is motivated by the fact that falsification engines aim to find a violation of $\phi$, and hence, the encoding above forces the control signal to be within the ranges of interest.

Next, \toolname~uses these abstractions to guide a system falsification engine. System falsification refers to a set of tools that take as input a dynamical system $f$ and an STL formula $\phi$ and find a physical-level trajectory $\psi_x = \{x_t\}_{t = 0}^H$ that violates $\phi$, i.e.,
\begin{align}
    \overline{\psi}_x = \texttt{Falsify}(f,\phi). 
\end{align}

If $\overline{\psi}_x$ is not empty, then the computed trajectory $\psi_x$ is guaranteed to violate the STL requirements $\phi$, i.e., $\psi_x \not \models \phi$. Note that the bar in $\overline{\psi}_x$ is used to emphasize the fact that $\overline{\psi}_x$ is not a \emph{concrete} vulnerability since it was computed based on the abstraction of cyber trajectories using control signal ranges (Lines 11-12, Algorithm~\ref{alg:rampo}).

Since the abstraction of cyber trajectories using range of control signals is a sound approximation, then whenever $\overline{\psi}_x$ is empty, one can conclude that the system is free from all cyber-kinetic vulnerabilities. On the other hand, if $\overline{\psi}_x$ is not empty, then our tool will extract the cyber trajectory corresponding to $\overline{\psi}_x$ as follows:
\begin{align}
\overline{\psi}_p &= \texttt{Extract-Path}(\overline{\psi}_x) \nonumber \\
&= \Big\{  \texttt{PATH}(h(\texttt{Element}(\overline{\psi_x}, t))) \Big\}_{t = 0}^H.    
\end{align}
We refer to $\overline{\psi}_p$ as an abstract vulnerability (Lines 13-16, Algorithm~\ref{alg:rampo}).

Since the abstract vulnerability $\overline{\psi}_p$ is extracted from a physical trajectory that violated the safety requirements, it is likely a \emph{concrete} vulnerability may exist in the system. To that end, \toolname~will explore $\overline{\psi}_p$ by iteratively refining the control signal ranges around $\overline{\psi}_p$.
In particular, \toolname~provides two different strategies for the iterative abstraction refinement of $\overline{\psi}_p$ that are explained in detail in Section~\ref{sec:CEGAR} and Algorithm~\ref{alg:rampo}. These abstraction refinement approaches can either find \emph{concrete} cyber vulnerabilities $\psi_p$, find other abstract cyber trajectories $\overline{\psi}^*_p$ that need to be subsequently refined, or certify that a set of abstract trajectories $\underline{\psi}_p$ will not lead to a vulnerability (Lines 19-23, Algorithm~\ref{alg:rampo}).

\begin{algorithm}[t]
 \caption{\toolname}
 \label{alg:rampo}
 \begin{algorithmic}[1]
 \renewcommand{\algorithmicrequire}{\textbf{Input:}}
 \renewcommand{\algorithmicensure}{\textbf{Output:}}
 \REQUIRE Safety specification $\varphi$, System dynamics $f$, binary-level control program $g$
 \ENSURE  Set of Cyber-Kinetic Vulnerabilities $\Psi_p^\varphi$
  \STATE Initialize $\Psi_p^\varphi:= \texttt{Empty Set}$ 
  \STATE Initialize $\overline{\Psi}_p:= \texttt{Empty Set}$ 
  \STATE Initialize $\underline{\Psi}_p:= \texttt{Empty Set}$ 
  \STATE \textbf{\ul{Step 1: Extract path constraints \& ranges:}}
  \STATE $(c^{(1)}\!\!, g^{(1)}), .., (c^{(k)}\!\!, g^{(k)}) = \texttt{Symbolic-Execution}(g)$ \label{ln:symbolic}
  \FOR {Each path $p$ in $\{1,\ldots, k\}$} 
    \STATE $\overline{u}_p = \max_{y \in \mathbb{PF}^o} g^{(p)}(y) \; \text{ subject to } \; c^{(p)}(y) = \texttt{True}$
    \STATE $\underline{u}_p = \min_{y \in \mathbb{PF}^o} g^{(p)}(y) \; \text{ subject to } \; c^{(p)}(y) = \texttt{True}$
  \ENDFOR
  \STATE \textbf{\ul{Step 2: Abstract all trajectories \& falsify:}}
    \STATE $\lnot \phi \; \leftarrow \;$ Equation~\eqref{eq:phi_1}
    \STATE $\overline{\psi}_x = \texttt{Falsify}(f, \phi)$
    \IF {$\texttt{Not-EMPTY}(\overline{\psi}_x)$}
        \STATE $\overline{\psi}_p = \texttt{Extract-Path}(\overline{\psi}_x)$
        \STATE $\overline{\Psi}_p = \overline{\Psi}_p \cup \{ \overline{\psi}_p \}$
    \ENDIF
    \WHILE{\texttt{Not-EMPTY}($\overline{\Psi}_p$)}
        \STATE \textbf{\ul{Step 3: Refine abstract trajectories:}}
        \STATE Pop one $\overline{\psi}_p$ from $\overline{\Psi}_p$
        \STATE $(\psi_p, \overline{\psi}^*_p, \underline{\psi}_p) = \texttt{Abstraction-Refinment}(f, \varphi, \overline{\psi}_p)$
        \STATE $\Psi_p^\varphi = \Psi_p^\varphi \cup \psi_p$
        \STATE $\overline{\Psi}_p = \overline{\Psi}_p \cup \{ \overline{\psi}^*_p \}$
        \STATE $\underline{\Psi}_p = \underline{\Psi}_p \cup \{ \underline{\psi}_p \}$
        \STATE \textbf{\ul{Step 4: Explore unexplored abstract trajectories:}}
        \STATE $\lnot \phi \; \leftarrow \;$ Equation~\eqref{eq:phi_2}
        \STATE $\overline{\psi}_x = \texttt{Falsify}(f, \phi)$
        \IF {$\texttt{Not-EMPTY}(\overline{\psi}_x)$}
            \STATE $\overline{\psi}_p = \texttt{Extract-Path}(\overline{\psi}_x)$
            \STATE $\overline{\Psi}_p = \overline{\Psi}_p \cup \{ \overline{\psi}_p \}$
        \ENDIF 
    \ENDWHILE
 \RETURN $\Psi_p^\varphi$
 \end{algorithmic} 
 \end{algorithm}
 
Finally, \toolname~will search for new abstract vulnerabilities over the unexplored abstract trajectories. This is achieved by forcing the falsification engine to search over the control signal ranges that are not considered yet by encoding them in the STL formula as:
\begin{align}
\lnot\phi = \lnot\varphi \land \Big[ \bigwedge_{\psi_p \in \underline{\Psi}_p} \bigwedge_{t = 0}^H 
&\lnot \left(
\texttt{Element}(\underline{\texttt{Range}}(\psi_p),t)
\le u_t \right) \Big] \nonumber \\
\land \Big[ \bigwedge_{\psi_p \in \underline{\Psi}_p} \bigwedge_{t = 0}^H 
&\lnot \left(
u_t \le \texttt{Element}(\overline{\texttt{Range}}(\psi_p), t ) \right) \Big],
\label{eq:phi_2}
\end{align}
where $\varphi$ is the original safety requirements and $\underline{\Psi}_p$ is the set of the abstract trajectories explored so far (Lines 25-30, Algorithm~\ref{alg:rampo}).

By refining the abstractions of the explored cyber vulnerabilities and searching over unexplored abstract trajectories, \toolname~will iteratively identify new abstract cyber trajectories and search for concrete vulnerabilities within those abstract cyber trajectories until all cyber trajectories are covered. This process is summarized in Algorithm~\ref{alg:rampo} and Figure~\ref{fig:framework_overview}.

\subsection{Correctness of Algorithm~\ref{alg:rampo}}
The soundness and completeness of Algorithm~\ref{alg:rampo} relies on the soundness and completeness of the falsification engine (\texttt{Falsify}). This follows from the fact that \toolname~uses sound abstraction of the cyber trajectories and relies on the falsification engine to discard abstract trajectories when no abstract physical-level trajectory violates the safety requirements.
Unfortunately, off-the-shelf falsification engines rely on stochastic optimization to reason about non-convex constraints, and hence, these falsification engines are sound but not complete, rendering \toolname~to be sound but not complete.

\subsection{Extension to Cyber-Physical-Kinetic Vulnerabilities}
Algorithm~\ref{alg:rampo} can be directly extended to enumerate cyber-physical-kinetic vulnerabilities. As captured by Problem~\ref{prob:CPS_kinetic}, the main difference between cyber-kinetic vulnerabilities and cyber-physical-kinetic vulnerabilities is the need to find additional sensor manipulation signals $\{s_t\}_{t = 0}^H$ that can lead to violation of safety requirements. To that end, one can treat the signal $s$ as an additional input signal to the falsification engine and rely on the falsification engine to simultaneously search for $\psi_x$ and $\psi_s$. An example of this extension is provided in the numerical examples in Section~\ref{sec:exp3}.

\section{Abstraction Refinement of Abstract Vulnerabilities}
\label{sec:CEGAR}

As shown in Algorithm~\ref{alg:rampo}, whenever an abstract vulnerability $\overline{\psi}_p$ is found, \toolname~needs to refine this abstraction to identify concrete vulnerabilities, find other potential abstract vulnerabilities, or exclude some abstract trajectories from the search space. In this section, we provide two alternative approaches for the iterative abstraction refinement.

\begin{figure}
    \centering
    \includegraphics[width=\linewidth]{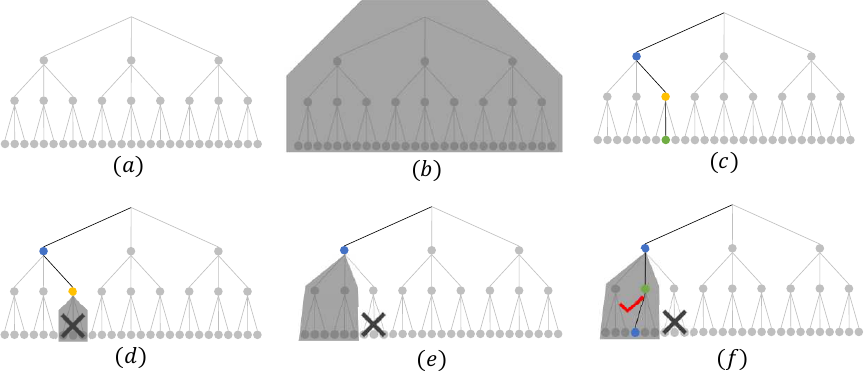}
    \vspace{-3mm}
    \caption{A pictorial representation of the Counter-Example Guided Abstraction Refinement process. (a) A tree of all possible cyber trajectories for a controller code with 3 paths and $H = 3$. (b) In the first iteration, \toolname \ will abstract all the trajectories in the search tree by considering the control signals that can be produced by all trajectories. (c) The falsification engine finds a physical-level trajectory that violates the safety requirements and \toolname \ extracts the corresponding trajectory. (d) The control signal ranges are refined/concretized around the trajectory found from the previous step except for the last time step. The falsification engine could not find a corresponding physical-level trajectory that violates the requirements, and hence, this sub-tree is removed from the search space. (e) \toolname \ backtracks one level up and abstracts the cyber trajectories around the truncated trajectory found before. (f) The falsification engine reported a physical-level vulnerability in the search tree, and the vulnerability was checked to be a concrete one.}
    \label{fig:CEGAR}
\end{figure}

\subsection{Linear-Search Based Refinement}
In the first abstraction refinement approach, given an abstract vulnerability $\overline{\psi}_p$ of length $l \le H$, our tool concretizes/refines the ranges around this specific $\overline{\psi}_p$. This can be encoded in the STL formula $\phi$ as follows:
\vspace{-2mm}
\begin{align}
    \lnot\phi = \lnot\varphi & \land \Bigg[ \bigwedge_{t = 0}^{l}
\texttt{Element}(\underline{\texttt{Range}}(\psi_p),t)
\le u_t \Bigg] \nonumber \\
& \land \Bigg[ \bigwedge_{t = 0}^{l} 
u_t \le 
\texttt{Element}(\overline{\texttt{Range}}(\psi_p),t)
\Bigg].
\label{eq:phi_3}
\end{align}
The falsification engine then uses $\phi$ to search for a trajectory of the physical system. Note that even if a trajectory is found, this does not account for a concrete vulnerability since the behavior of the control program $g$ is still abstracted as a range of control signals (recall the two levels of abstractions in Section 4.1). This requires another call to the Falsification engine, this time using the physical model $f$ augmented with the control program $g$ constrained to the path constraints $c^{(\texttt{Element}(\overline{\psi}_p, 1))}, \ldots, c^{(\texttt{Element}(\overline{\psi}_p, l))}$. If a concrete vulnerability $\psi_x$ was found, then it is added to the set of concrete cyber-kinetic vulnerabilities. Regardless of whether a concrete vulnerability is found, the abstract trajectory $\overline{\psi}_p$ is added to the set of explored trajectories $\underline{\psi}_p$ (Lines 2-11, Algorithm~\ref{alg:linear}).

If a concrete vulnerability is not found by the process above, we iteratively reduce the abstraction refinement by removing the range constraints in $\phi$ (as pictorially illustrated in Figure~\ref{fig:CEGAR}). That is, starting from $i = 1$ until $i = l$, we will iteratively call the falsification engine to find a trajectory of the physical system that violates the safety requirements by encoding these range constraints in $\phi$ as follows:
\vspace{-2mm}
\begin{align}
    \lnot\phi = \lnot\varphi & \land \Bigg[ \bigwedge_{t = 0}^{l-i}  
\texttt{Element}(\underline{\texttt{Range}}(\psi_p),t)
\le u_t \Bigg] \nonumber \\
& \land \Bigg[ \bigwedge_{t = 0}^{l-i} 
u_t \le 
\texttt{Element}(\overline{\texttt{Range}}(\psi_p),t)
\Bigg].
\label{eq:phi_4}
\end{align}
Whenever the falsification engine fails to find a violation of $\phi$ (for $i = 1, \ldots, l$), the truncated version of $\overline{\psi}_p$ (i.e., $\{p_t\}_{t = 0}^i$) is added to the set of explored trajectories $\underline{\psi}_p$. Similarly, whenever the falsification engine finds an abstract vulnerability, the newly discovered abstract vulnerability is added to $\overline{\psi}^*_p$. This process is summarized in~Figure~\ref{fig:CEGAR} and Algorithm~\ref{alg:linear}.

\begin{algorithm}[t]
 \caption{$\texttt{Abstraction-Refinement-Linear}$}
 \label{alg:linear}
 \begin{algorithmic}[1]
 \renewcommand{\algorithmicrequire}{\textbf{Input:}}
 \renewcommand{\algorithmicensure}{\textbf{Output:}}
 \REQUIRE Safety requirements $\varphi$, Physical System Model $f$, Abstract cyber trajectory $\overline{\psi}_p$
 \ENSURE  Set of concrete cyber vulnerabilities $\psi_p$, 
        \\ Set of unexplored abstract vulnerabilities $\overline{\psi}^*_p$, 
        \\Set of explored abstract trajectories $\underline{\psi}_p$
  \STATE Initialize $\psi_p, \overline{\psi}^*_p, \underline{\psi}_p := \texttt{Empty Set}$  
  \STATE \textbf{Step 1: Search for concrete vulnerability}
  \STATE $\lnot \phi \; \leftarrow \; $ Equation~\eqref{eq:phi_3}
  \STATE $\overline{\psi}_x = \texttt{Falsify}(f, \phi)$
    \IF {$\texttt{Not-EMPTY}(\overline{\psi}_x)$}
        \STATE $f_{\text{cl}} = \texttt{Form-Closed-Loop-Model}(f, g)$
        \STATE $\psi_x = \texttt{Falsify}(f_{\text{cl}}, \phi)$
        \IF {$\texttt{Not-EMPTY}(\psi_x)$}
            \STATE $\psi_p = \overline{\psi}_p$
            \STATE $\underline{\psi}_p = \underline{\psi}_p \cup \overline{\psi}_p$
        \ENDIF
    \ELSE
        \FOR{$i = 1$ to $l$}
            \STATE $\lnot \phi \; \leftarrow \; $ Equation~\eqref{eq:phi_4}
            \STATE $\overline{\psi}_x = \texttt{Falsify}(f, \phi)$
            \IF {$\texttt{Not-EMPTY}(\overline{\psi}_x)$}
                \STATE $\overline{\psi}^*_p = \overline{\psi}^*_p \cup \left\{\{ \texttt{Element}(\psi_p,t) \}_{t = 0}^{l-i} \right\}$
                \STATE Exit the FOR loop
            \ELSE
                \STATE $\underline{\psi}^*_p = \underline{\psi}^*_p \cup \left\{\{ \texttt{Element}(\psi_p,t) \}_{t = 0}^{l-i} \right\}$
            \ENDIF
        \ENDFOR
    \ENDIF
 \RETURN $\psi_p, \overline{\psi}^*_p, \underline{\psi}_p$
 \end{algorithmic} 
 \end{algorithm}

\subsection{Binary-Search Based Refinement}

While Algorithm~\ref{alg:linear} relaxes the range constraints along $\overline{\psi}_p$ in a linear fashion, our second approach for abstraction refinement is to relax the range constraints in a binary-search fashion. That is, we replace the loop in Step 14 of  Algorithm~\ref{alg:linear} with one that starts with $i = l/2$ and then selects the next value of $i$ based on the success/failure of finding an abstract vulnerability. Compared to the linear-search-based approach, this binary search is more suitable for traversing trees with high values of $l$ (hence the height of the tree being explored by the abstraction refinement). 

%% file: sections/6_experiment_result.tex
\section{Experimental Evaluation}
\label{sec:experiment}

We implemented \toolname \ in Python where we used \emph{angr}~\cite{angr} for symbolic code execution, \emph{Z3}~\cite{z3} to compute the ranges of the control signal in each path, and \emph{S-TaLiRo}~\cite{s-taliro} as the falsification engine. In this section, we perform a series of numerical analyses to evaluate the effectiveness and the scalability of our tool. First, we conduct a series of studies to compare against state-of-the-art falsification of closed-loop systems and to study the effect of varying different parameters (e.g., the complexity of the physical system and the complexity of the cyber system). We utilize two metrics for this study: (i) the execution time and (ii) the number of identified cyber-kinetic vulnerabilities. Next, we will study the ability to identify both cyber-kinetic and cyber-physical-kinetic attacks using a sophisticated model for engine control systems.

\begin{figure*}[t]
    \centering
    \includegraphics[width=\linewidth]{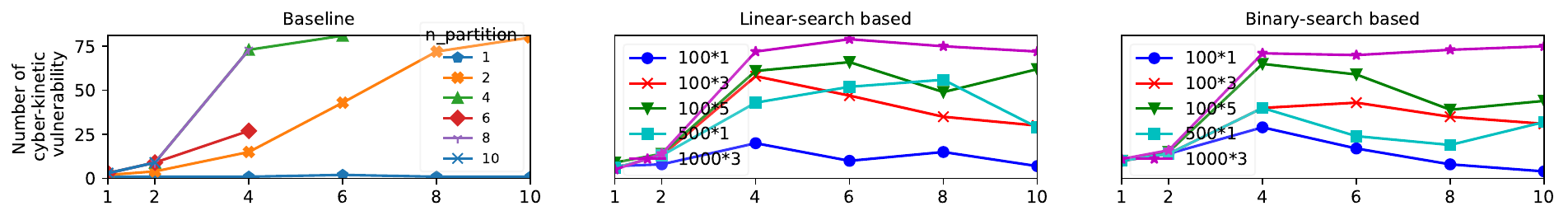}
    \includegraphics[width=\linewidth]{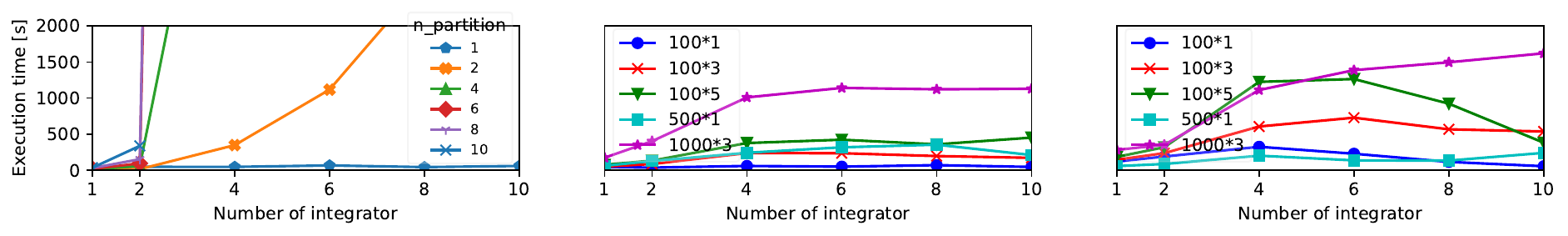}
    \vspace{-8mm}
    \caption{The number of cyber-kinetic vulnerabilities found and the execution time. $n\_partition$ refers to the number of partitions among each dimension of the state space in the model. 
    In the brute-force (baseline) approach, as the number of partitions increases, the execution time increases exponentially. 
    For both the linear- and binary-search-based abstraction refinement approaches of \toolname, the execution time is almost constant as the number of integrators in the model increases. This 
    leads to a speedup that ranges from $3\times$ to $98\times$ while computing the same number of vulnerabilities.
    } 
    \label{fig:unique_path}
\end{figure*}

\subsection{Experiment 1: Comparison Against Black-Box Falsification}
We start by comparing \toolname \ to \emph{S-TaLiRo}, one of the state-of-the-art CPS falsification tools. While we use \emph{S-TaLiRo} internally within \toolname \ to find falsifying trajectories of the open-loop system $f$ (up to specified path constraints), \emph{S-TaLiRo} can also be used to falsify the entire closed-loop system (the dynamical system $f$ along with the controller $g$). In this experiment, we show that analyzing the controller $g$ away from the dynamical system $f$ (using our approach) will lead to identifying a more significant number of vulnerabilities in the system compared to the as-is use of \emph{S-TaLiRo}. 

To that end, we consider a model of a drone moving vertically. We use a
simple double integrator dynamical system $f$ to capture the vertical movement of the drone. The safety requirement is for the drone's position to be always below $0.98$. We also consider the following control program, which implements a gain scheduling controller:
\begin{lstlisting}[caption={Control program for Experiments 1-3}, label={lst:drone_code}]
double control(double x1){
  if (x1 < -1) x1 = -1;
  if (x1 > 1) x1 = 1;

  if (-1 <= x1 && x1 <= -0.5){
    u = 4 * x1 - 6;
  }else if(-0.5 < x1 && x1 < 0.5){
    u = 2 * x1 + 9;
  }else if(0.5 <= x1 && x1 <= 1){
    u = -4 * x1 - 6;
  }
  return u;
}
\end{lstlisting}

Note that \emph{S-TaLiRo} aims to find only one falsifying trajectory while our aim in Problem~\ref{prob:cyber_kinetic} is to enumerate all possible cyber-kinetic vulnerabilities. To that end, we ran \emph{S-TaLiRo} 50 times with random seeds to force it to explore different vulnerabilities. We ran \toolname \ while restricting the number of calls to the \texttt{Falsify} function to 50. Our experiments show that \toolname \ was able to identify 14 cyber-kinetic vulnerabilities while S-TaLiRo-alone was able to find only 9 of those vulnerabilities. The execution time of \toolname \ was $191$ seconds, while for S-TaLiRo-alone was $147$ seconds. This result shows one of the main benefits of combining falsification and code analysis tools. By splitting the analysis between the two domains (cyber and physical domains), one can achieve better coverage of the different cyber trajectories in the system.

\subsection{Experiment 2: Scalability with respect to complexity of the physical system model}
Our second experiment aims to evaluate the scalability of our tool with respect to the number of states in the dynamical model $f$. To that end, we use the same control program in Listing~\ref{lst:drone_code}, and we vary the number of integrators in the system from $1$ to $10$.

To obtain a baseline for comparison, we perform an exhaustive search by partitioning the space of initial states (used to find falsifying trajectories by \emph{S-TaLiRo}) into hypercubes. By constraining \emph{S-TaLiRo} to find falsifying trajectories that start from different initial states, we can increase the number of identified vulnerabilities. Nevertheless, and as shown in Figure~\ref{fig:unique_path} (left), this brute force search scales poorly as the number of states increases and as the granularity of the partitioning increases.

On the other hand, as shown in Figure~\ref{fig:unique_path} (middle and right), our tool with the linear- and binary-search-based abstraction refinement shows a similar trend to the baseline in terms of their ability to identify cyber-kinetic vulnerabilities while avoiding the exponential increase in the execution time. In general, the execution time of \toolname \ remained almost constant as the number of states increased. We repeated the same experiment multiple times while changing some of the internal parameters to the \texttt{Falsify} function on \toolname \ (which as explained above, is implemented using \emph{S-TaLiRo} but with open-loop dynamics and path constraints). In particular, different plots in Figure~\ref{fig:unique_path} show different settings where the number before the multiplication operator indicates the number of optimization steps taken in one falsification, and the latter numbers are the number of calls to \emph{S-TaLiRo} per one call to the \texttt{Falsify} function. 

For the same number of identified vulnerabilities, \toolname \ achieved a speedup that ranges from $3\times$ to $98\times$.

\begin{figure}[!t]
    \centering
    \includegraphics[width=\linewidth]{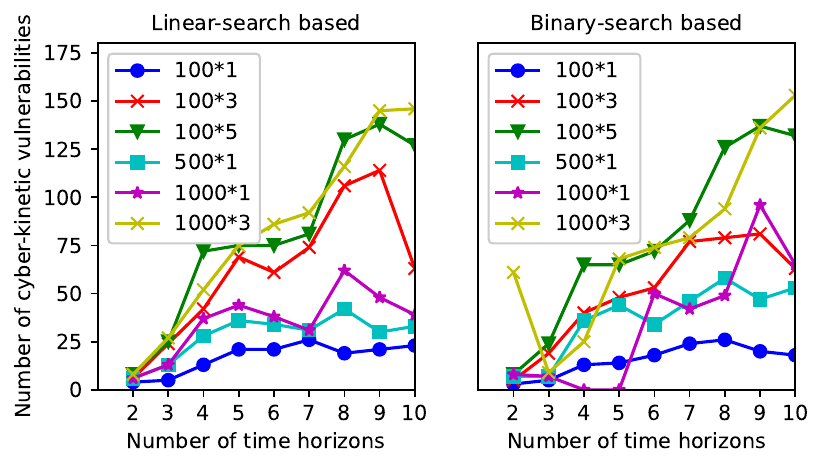}\\
    \includegraphics[width=\linewidth]{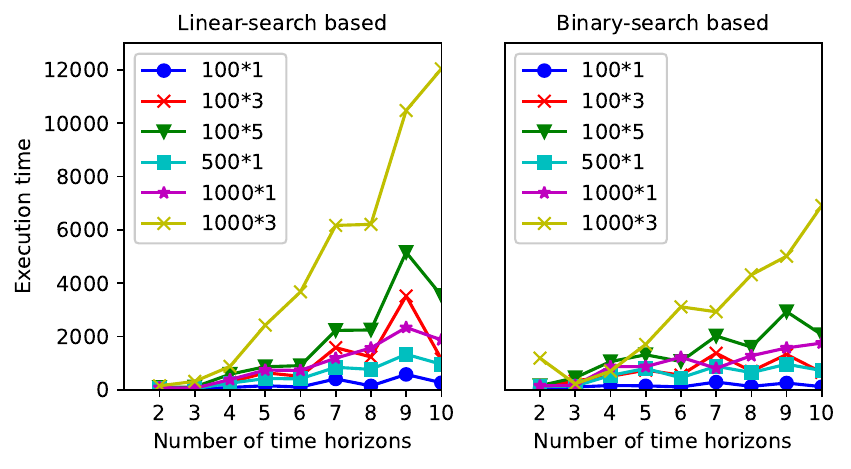}
    \caption{Number of cyber-kinetic vulnerabilities found and execution times in different lengths of time horizons}
    \label{fig:timehorizon_n_path}
\end{figure}

\subsection{Experiment 3: Scalability with respect to the complexity of the control software}
In this experiment, we study the scalability of \toolname \ as we increase the length of the cyber trajectories $H$. Recall that increasing $H$ results in an exponential growth in the number of cyber trajectories that need to be explored. Figure \ref{fig:timehorizon_n_path} (top) shows the number of identified vulnerabilities as $H$ increases for both the linear- and binary-search-based abstraction refinement approaches. As expected, the number of vulnerabilities increases as $H$ increases. Moreover, the number of vulnerabilities found by the linear- and binary-search-based abstraction refinement approaches are comparable among the different settings for the \texttt{Falsify} function.

On the other hand, \ref{fig:timehorizon_n_path} (bottom) shows the execution time for both the linear- and binary-search-based abstraction refinement approaches. As expected, the benefits of the binary search start to grow as the depth $H$ increases, resulting in an average of $2\times$ speedup for the binary-search-based abstraction refinement compared to the linear-search-based abstraction refinement approach. Moreover, and most importantly, the execution time of both the linear- and binary-search-based abstraction refinement approaches does \emph{not} increase exponentially with $H$, albeit with the exponential growth in the number of cyber trajectories. We hypothesize that this pattern is due to the CEGAR approach used by \toolname \, which can reason about several cyber trajectories simultaneously, which harnesses the exponential growth in the number of cyber trajectories.

\subsection{Experiment 4: Enumerating Cyber-Kinetic and Cyber-Physical-Kinetic Vulnerabilities}
\label{sec:exp3}
\begin{figure}[t]
    \centering
    \includegraphics[width=\linewidth]{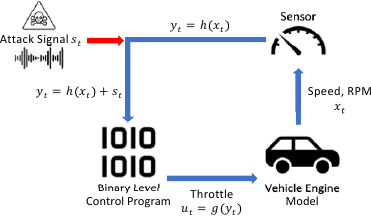}
    \caption{Vehicle engine model}
    \label{fig:engine}
\end{figure}

In this experiment, we evaluate the ability of \toolname \ to identify both cyber-kinetic and cyber-physical-kinetic vulnerabilities in complex systems. To that end, we use a model for a vehicle control signal (shown in Figure~\ref{fig:engine}). The model has two inputs, $Speed$ and $RPM$, and the one output, $Throttle$. The control program for this model is shown in Listing \ref{lst:engine_code} which has a total of $k = 4$ paths.
The safety requirement $\varphi$ is ``Either the $Speed$ or the $RPM$ is below a specified threshold.''

\begin{lstlisting}[caption={Control program for the vehicle engine model},label={lst:engine_code}]
double control(double RPM, double Speed){
  if (RPM > 3300){
    if (Speed > 80){
      Throttle = -RPM*0.002 - Speed*1.1 + 183.0;
    } else {
      Throttle = - RPM*0.001 + Speed*0.6 + 19.0;
    }
  } else {
    if (Speed > 80){
      Throttle = RPM*0.001 - Speed*1.7 + 216.0;
    } else {
      Throttle = - RPM*0.001 - Speed*0.6 + 139.0;
    }
  }
  return Throttle;
}
\end{lstlisting}

\begin{figure}
    \centering
    \includegraphics[width=\linewidth]{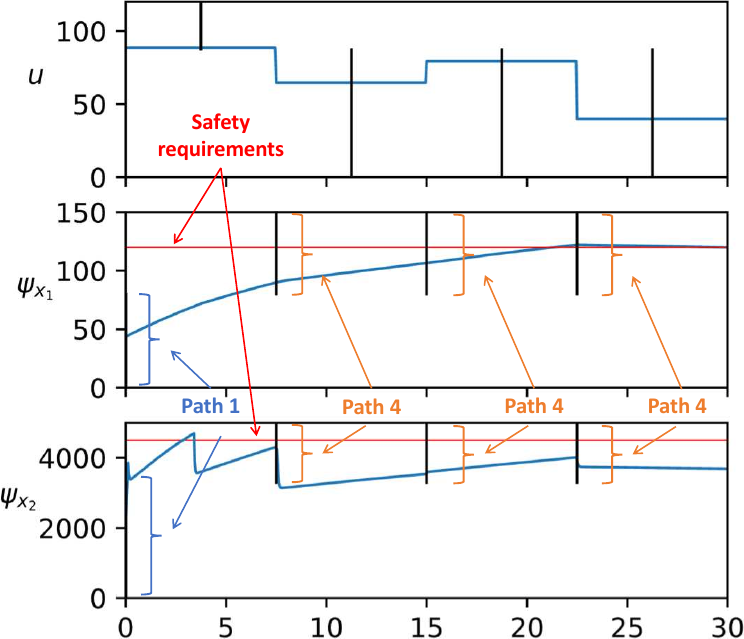}
    \caption{Cyber-kinetic vulnerability found in the vehicle engine model}
    \label{fig:without_injection}
\end{figure}

We started by searching for cyber-kinetic vulnerabilities using \toolname. Figure \ref{fig:without_injection} shows the only cyber-kinetic vulnerability found by the binary-search-based abstraction refinement approach. This vulnerability corresponds to the cyber trajectory $(p_0 = 1, p_1 = 4, p_2 = 4, p_3 = 4)$. The corresponding trajectories of $\psi_{x_1}$ and $\psi_{x_2}$ are shown in Figure \ref{fig:without_injection} where cross the bars that indicate corresponding paths.   

Next, we examine whether a small amount of physical attack signals can lead to \emph{other} kinetic attacks, i.e., can a small amount of physical attack signal force the control program to go through a different cyber trajectory that can lead to a new kinetic attack? To that end, we restrict the physical attack signal $\psi_s$ to be $ \leq 2.0\%$ of the original sensor signals ($Speed$ and $RPM$). We considered three different configurations. In the first one, the attacker can change only $Speed$, in the second one, the attacker can only change $RPM$, while in the last configuration, the attacker can affect both $Speed$ and $RPM$.

\begin{figure}
    \centering
    \includegraphics[width=\linewidth]{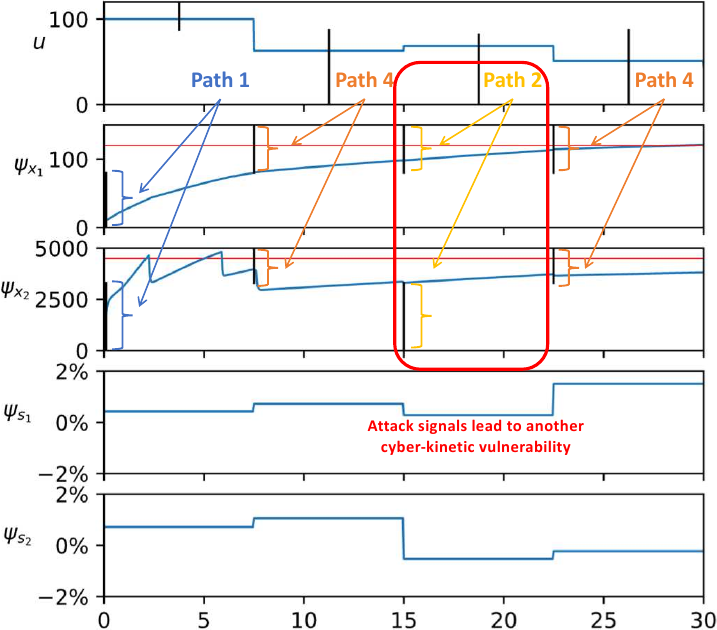}
    \caption{Another cyber-kinetic vulnerability found only when attack signals are injected}
    \label{fig:with_injection}
\end{figure}

For the first two settings (attacking either $Speed$ alone or $RPM$ alone), our tool was not able to find any \emph{new} cyber trajectory that can lead to vulnerability (other than the one shown in Figure~\ref{fig:without_injection}). On the other hand, by allowing the attacker to corrupt both the $Speed$ and $RPM$, our tool identified another cyber trajectory that $(p_0 = 1, p_1 = 4, p_2 = 2, p_3 = 4)$ that can lead to vulnerabilities. The cyber-physical-kinetic vulnerability is shown in Figure \ref{fig:with_injection}.

By analyzing the third time step in Figure~\ref{fig:with_injection}, one can notice that the $\psi_{x_2}$ trajectory is close to violating the constraints of $p = 2$. That is, a small perturbation in the sensor signal (either intentional by an attacker or by noise) can force the code to execute a different path in the control program $g$. This shows the effectiveness of our tool to identify these corner cases, which can be used to fix the control program to be more robust and secure.

%% file: sections/7_conclusion.tex
\section{Conclusion}
\label{sec:conclusion}
This paper proposed a novel tool, named \toolname, that can perform binary code analysis to identify cyber kinetic vulnerabilities in CPS.
Our tool analyzes the binary level control program to extract the ranges of control signal inputs to the physical system. It uses this information to abstract the code and guide falsification tools to identify possible kinetic vulnerabilities. The tool iteratively refines the abstraction until concrete cyber-kinetic vulnerabilities are found or the entire space of cyber trajectories is covered. We provided different approaches for the abstraction refinement process and generalized the framework to find cyber-physical-kinetic vulnerabilities. Our numerical analysis shows that our tools scale more favorably compared to off-the-shelf tools and are able to harness the exponential growth in the search space when the temporal evolution of the code is considered leading to $3\times$-$98\times$ speed up in execution time while identifying the same number of vulnerabilities.